\documentclass[aps,prl,showpacs,preprintnumbers,twocolumn,superscriptaddress]{revtex4-2}
\usepackage{amsmath,amssymb}
\usepackage{bm}
\usepackage{tipa}
\usepackage{upgreek}
\usepackage{comment}
\usepackage{mathrsfs}
\usepackage{graphicx}
\usepackage{braket}
\usepackage{enumitem}
\usepackage{mathbbol}
\usepackage{booktabs}
\usepackage{gensymb}
\usepackage[normalem]{ulem}
\usepackage{color}
\usepackage[colorlinks,bookmarks=true,citecolor=blue,linkcolor=red,urlcolor=blue]{hyperref}
\usepackage{hyperref}
\usepackage{pifont}
\usepackage{dsfont}

\def \r{{\mathbf r}}

\def \k{{\mathbf k}}
\def \q{{\mathbf q}}
\def \Q{{\mathbf Q}}

\def \0{{\mathbf 0}}

\begin{document}

  \title{Nodal d-wave pairing from spin fluctuations in a thermally disordered anti-ferromagnet}
	\author{Nick Bultinck}
	\affiliation{Department of Physics, Ghent University, Krijgslaan 281, 9000 Gent, Belgium}

\begin{abstract}
   We consider electron pairing in a two-dimensional thermally disordered itinerant anti-ferromagnet. It is shown that transverse spin fluctuations in such a state can give rise to superconductivity with a sizeable critical temperature $T_c$. Below $T_c$ there is quasi-long-range spin-singlet and $d_{x^2-y^2}$ superconducting order, together with fluctuating triplet order at momentum $(\pi,\pi)$. The singlet pairs we find are tightly bound together, and the pair wavefunction has a purely inter-sublattice structure due to the U$(1)$ spin rotation symmetry of the anti-ferromagnet.
\end{abstract}

 	\maketitle

\emph{Introduction -- }It is well-known that anti-ferromagnetic (AFM) spin fluctuations generate an attractive interaction between electrons which favors pairing in the spin-singlet $d_{x^2-y^2}$-channel \cite{Hirsch1985,Emery1986,Scalapino1986,Miyake1986}. In the original approach for connecting AFM fluctuations and $d$-wave superconductivity \cite{Hirsch1985,Emery1986,Scalapino1986,Miyake1986,Scalapino1987,Lee1987,Monthoux1991,Monthoux1992,Scalapino1994,Scalapino2012}, the normal state is a conventional Fermi liquid. In the underdoped cuprates, however, the normal state does not seem to fit the standard Fermi liquid mould. For example, one of the intriguing properties of the underdoped cuprate normal state is the pseudogap phenomenon, manifested in the form of a suppressed density of states at the Fermi energy and Fermi arcs in photoemission spectra. Motivated by these experimental observations, we consider a magnetic pseudogap state in the form of a thermally disordered AFM --i.e. an AFM above its critical temperature (which is $T=0$), but well below the mean-field transition temperature-- as the parent state for superconductivity. This magnetic pseudogap state has a suppressed density of states at the Fermi energy due to the fluctuating anti-ferromagnetism. It also produces a spectral weight consisting of $4$ small Fermi pockets centered at $(\pm \pi/2,\pm \pi/2)$ with faint backsides, resembling the Fermi arcs seen in photoemission. Inspired by the well-established connection between AFM spin fluctuations and $d$-wave pairing in Fermi liquids, our goal here is to investigate the role of spin fluctuations for mediating superconductivity in such a thermally disordered AFM. In particular, we focus on spin fluctuations which are the finite-temperature remnants of the AFM Goldstone modes.

To set up our calculations we start from the effective theory for a thermally disordered AFM presented in Ref. \cite{Vasiliou2023}. The effective action is derived from a simple mean-field and random phase approximation (RPA). However, one of its main advantages is that the effective theory contains scattering vertices between the electrons and Goldstone modes which are not entirely phenomenological. Instead these vertices are calculated starting from the microscopic Hamiltonian. And as the tendency of an interaction to promote $d$-wave pairing crucially relies on its structure in momentum space, having an explicit expression for the scattering vertices is key for our analysis. 

Within the effective theory we find that spin fluctuations indeed give rise to superconductivity with a critical temperature which is high compared to conventional phonon-driven superconductors. Crucially, due to the symmetry properties of the AFM, the gap function we find cannot be a featureless $s$-wave state but instead needs to have a non-trivial momentum dependence. A further important property of the effective theory is that the fermion fields are defined in a `rotating frame' \cite{Shraiman1988,Shraiman1990,Schulz1990,Schulz1994}, in order to ensure that long-wavelength Goldstone modes decouple from the electrons \cite{Vasiliou2023,Schrieffer1995}. It is also this rotating frame which ultimately ensures that the superconducting order parameter we find is spin-singlet. At the end of the manuscript we discuss a few predictions of our theory which can be used to test whether superconductivity in a particular model or material indeed originates from spin fluctuations in a thermally disordered AFM.

Our approach is closely related to both the spin-bag \cite{Schrieffer1988,Schrieffer1989} and spin-fermion \cite{Monthoux1991,Monthoux1992,Schrieffer1995,Chubukov1995,Altshuler1995,Chubukov1997,Schmalian1998,Abanov2003,Chubukov2008} models studied in the early days of high-$T_c$ superconductivity. However, it also differs from these pre-existing theories in some crucial ways. In particular, unlike in the spin-fermion model, the bare coupling between electrons and spin fluctuations in our effective theory cannot be described by a purely local and instantaneous interaction. Our approach also differs from the original spin-bag theory because we focus on transverse spin fluctuations, previously studied in \cite{Frenkel1990,Rowe2015,Romer2016}, and use a rotating frame which removes strong inter-band scattering terms \footnote{See Refs. \cite{Sedrakyan2010,Ye2019,Ye2023} for a complementary perspective. In these works, the strong inter-band scattering terms from transverse spin fluctuations are not eliminated via a rotating frame, but their effect on the electron Green's function is explicitly taken into account via an infinite sum of diagrams in an eikonal approximation.} and leads to a spin-singlet superconductor with nodes on the Fermi surface. Moreover, we consider spin fluctuations with a non-zero thermal mass to generate the effective attraction, and use the complete dynamical interaction to calculate $T_c$.

\begin{figure}
\includegraphics[scale=0.35]{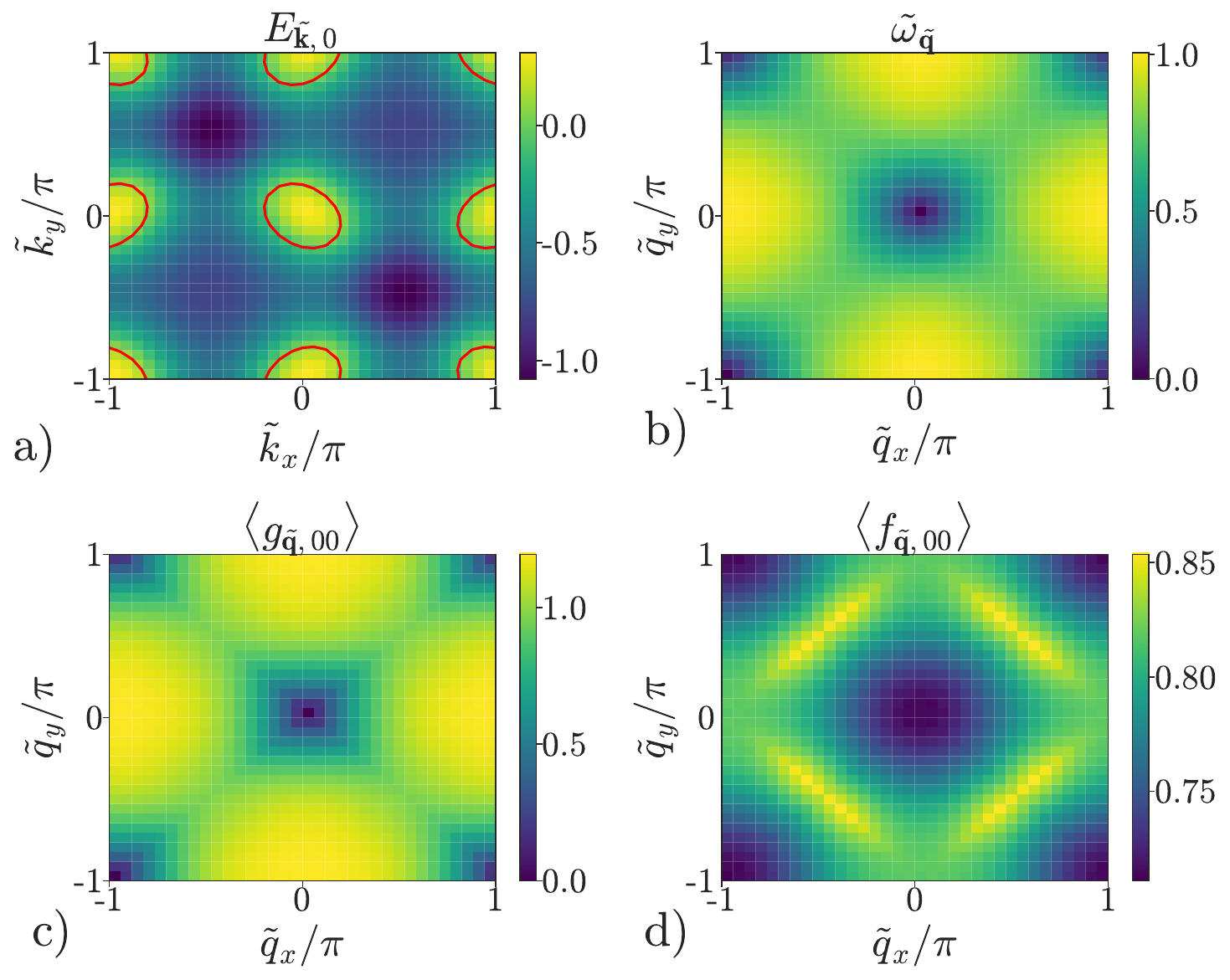}
\caption{(a) Mean-field valence band energy of the AFM insulator. The mean-field hybridization strength between spin up and down electrons is $\sim 1.93$. The Fermi surface obtained by $\sim 12\%$ hole doping is shown in red. (b) Goldstone mode dispersion relation $\tilde{\omega}_{\tilde{\q}}$. (c)-(d) Averaged absolute values of the vertices in Eq. \eqref{SelB} for electrons in the valence band ($\alpha=\beta=0)$ as a function of the pseudo-momentum transfer $\tilde{\q}$, where the averaging is over the incoming pseudo-momentum. Results are obtained on a $34\times 34$ pseudo-momentum grid.} \label{fig:SeffData}
\end{figure}

\emph{Model and results -- }As a microscopic model we consider the Hubbard model on the square lattice:
\begin{equation}\label{Hubbard}
    H = -t \sum_{\langle ij\rangle}\sum_\sigma c^\dagger_{i,\sigma} c_{j,\sigma} -t'\sum_{\langle\langle ij\rangle\rangle} c^\dagger_{i,\sigma} c_{j,\sigma} + h.c. + U\sum_i n_{i,\uparrow} n_{i,\downarrow}\,, 
\end{equation}
where the first (second) sum is over nearest (next nearest) neighbors. In the third term $n_{i,\sigma}$ is the number of electrons with spin $\sigma$ on site $i$. We take $t\equiv 1$, $t' = -0.35$, and $U=5$. At half filling, the system is an insulating antiferromagnet. We assume that the AFM moments are ordered in the $x$-direction, and work with following imaginary-time effective action to describe the interacting electrons and AFM Goldstone modes at mean-field+RPA level: 
\begin{equation}\label{Seff}
S_{\rm eff} = S_{el} + S_V + S_B + S_{el-B}
\end{equation}
The first term is the kinetic energy term for the fermions
\begin{equation}\label{Sel}
S_{el} = \int\mathrm{d}\tau \sum_{\tilde{\k}} \sum_\alpha \bar{\psi}_{\tilde{\k},\alpha} (\partial_\tau + E_{\tilde{\k},\alpha} -\mu )\psi_{\tilde{\k},\alpha}\,,
\end{equation}
where $\alpha=0,1$ labels the mean-field bands of the AFM insulator -- $0$ ($1$) is the valence (conduction) band. The mean-field valence band energy $E_{\tilde{\k},0}$ is shown in Fig. \ref{fig:SeffData} (a). The conduction band is separated by a gap of order $U$. The AFM breaks translation over one site, but is invariant under the action of $T'_{\r} = e^{i\r\cdot \Q\, \sigma^z/2} T_{\r}$, where $\Q = (\pi,\pi)$, $\sigma^i$ are the Pauli spin matrices, and $T_\r$ implements a translation by lattice vector $\r$. The pseudo-momenta $\tilde{\k}$ are the conserved quantum numbers (modulo reciprocal lattice vectors) associated with the $T'_\r$ symmetry. The relation between states with pseudo-momentum $\tilde{\k}$ and crystal momentum $\k$ is given by:
\begin{equation}
f^\dagger_{\tilde{\k},\uparrow} := c^\dagger_{\tilde{\k}-\Q/2,\uparrow}\hspace{0.1 cm}\,,\hspace{0.5 cm}
f^\dagger_{\tilde{\k},\downarrow} := c^\dagger_{\tilde{\k}+\Q/2,\downarrow} \label{75}\,,
\end{equation}
where $f^\dagger_{\tilde{\k},\sigma}$ creates electrons with pseudo-momentum $\tilde{\k}$. 

Invariance under spin rotations around the $x$-axis implies that the effective theory is symmetric under $f^\dagger_{\tilde{\k},\uparrow}\leftrightarrow f^\dagger_{\tilde{\k}+\Q,\downarrow}$. This symmetry acts on the fermion fields $\bar{\psi}_{\tilde{\k},\alpha} = \sum_{\sigma=\uparrow,\downarrow}u(\tilde{\k})_{\alpha,\sigma}\bar{\psi}_{\tilde{\k},\sigma}$, where $u(\tilde{\k})_{\alpha,\sigma}$ are the coefficients of the mean-field single-particle states \cite{supplement}, as $\bar{\psi}_{\tilde{\k}+\Q,\alpha} \leftrightarrow  \pm \bar{\psi}_{\tilde{\k},\alpha}$. The signs depend on a choice of real gauge for $u(\tilde{\k})_{\alpha,\sigma}$. Here we use a gauge where all signs are positive. 

The second term $S_V$ in Eq. \eqref{Seff} contains the instantaneous two-body interaction for the electrons, which consists of the microscopic Hubbard interaction, and an interaction which is generated by integrating out the field conjugate to the Goldstone field $\phi(\r)$ \cite{Vasiliou2023}. The instantaneous interaction is written out in detail in the supplementary material \cite{supplement}. 

The dynamics of the Goldstone field is described by the third term in the effective action:
\begin{equation}\label{SB}
    S_B = \frac{1}{2}\int\mathrm{d}\tau\,\sum_{\tilde{\q}}\left( -\phi_{-\tilde{\q}}\partial^2_\tau \phi_{\tilde{\q}} + \omega_{\tilde{\q}}^2 \phi_{\tilde{\q}}\phi_{-\tilde{\q}}\right)\, ,
\end{equation}
where $\omega^2_{\tilde{\q}} = \tilde{\omega}_{\tilde{\q}}^2 + m^2$, with $\tilde{\omega}^2_{\tilde{\q}} \sim c^2\tilde{\q}^2$ near $\tilde{\q} = 0$, and $\tilde{\omega}^2_{\tilde{\q}} \sim c^2(\Q-\tilde{\q})^2$ near $\tilde{\q}=\Q$ ($c\approx 0.7)$. The complete Goldstone dispersion $\tilde{\omega}_{\tilde{\q}}$ is shown in Fig. \ref{fig:SeffData} (b). $m$ is a thermal mass for the Goldstone modes which takes into account that 2D AFM are disordered at non-zero temperatures. Below we will treat $m$ as a phenomenological parameter. The final term in Eq. \eqref{Seff} contains the electron-boson interaction
\begin{equation}\label{SelB}
\begin{split}
    S_{el-B} =&  \int\mathrm{d}\tau\, \frac{1}{\sqrt{N}}\sum_{\tilde{\q},\tilde{\k}}\bar{\psi}_{\tilde{\k},\alpha}\psi_{\tilde{\k}-\tilde{\q},\beta} \times \\
    & \left( g_{\tilde{\q},\alpha\beta}(\tilde{\k})\,\phi_{-\tilde{\q}} + f_{\tilde{\q},\alpha\beta}(\tilde{\k}) \,i\partial_\tau \phi_{-\tilde{\q}}\right)\,,
\end{split}
\end{equation}
where $N$ is the number of lattice sites. Following the procedure explained in Ref. \cite{Vasiliou2023}, the vertex functions $g$ and $f$ are constructed from solutions to the Bethe-Salpeter equation for the Hubbard model (which also determines the Goldstone mode spectrum $\tilde{\omega}_{\tilde{\q}}$). In Figs. \ref{fig:SeffData} (c-d) these vertices are shown for electron scattering within the valence band ($\alpha=\beta=0$). In particular, we show the absolute value of the vertices, averaged over the incoming pseudo-momentum, as a function of the pseudo-momentum transfer $\tilde{\q}$.

A final crucial property of the effective action is that the fermion fields are defined in a `rotating frame', which means that the bare/microscopic fermions $\bar{\psi}^b_{\r,\sigma}$ are related to the fields in Eq. \eqref{Sel} by $\bar{\psi}_{\r,\sigma} = \sum_{\sigma'}R_{\sigma\sigma'}(\tau,\r)\bar{\psi}^b_{\r,\sigma'}$. The $2\times 2$ matrix field $R(\tau,\r)$ is defined as
\begin{equation}
R(\tau,\r) = \exp\left(-i\left[\phi_z(\tau,\r)\sigma^z + \phi_y(\tau,\r)\sigma^y \right] \right)\,,
\end{equation}
where $\phi_z(\tau,\r)$ and $\phi_y(\tau,\r)$ contain momenta $\tilde{\k}$ in the first magnetic Brillouin zone, i.e. the Brillouin zone with reciprocal vectors $(\pi,\pm \pi)$, and are related to the Goldstone field in Eq.~\eqref{SB} by $\phi = \phi_z+(-1)^{\Q\cdot \r}\phi_y$. The reason for defining the fermion fields in a rotating frame is that these fermions decouple from the low-frequency and long-wavelength Goldstone modes \cite{Vasiliou2023,Schrieffer1995}, i.e. in this basis we have $\lim_{\tilde{\q}\rightarrow 0,\Q}g_{\tilde{\q},\alpha\beta}(\tilde{\k}) = 0$. 

We now proceed by changing the chemical potential $\mu$ to hole-dope the AFM. Upon changing the electron density one should actually redetermine the optimal mean-field state and the RPA collective modes that go into the construction of our effective action. However, for metallic systems calculating the scattering vertices $g$ and $f$ is challenging, as the Goldstone modes are partly hidden inside the particle-hole continuum. We will therefore keep the effective action as is, and simply change $\mu$. For our purposes we expect this to be a reasonable approximation at small doping -- in particular, we expect that the gross features in the momentum dependence of the vertex functions $g$ and $f$ will not change. 

\begin{figure}
\includegraphics[scale=0.31]{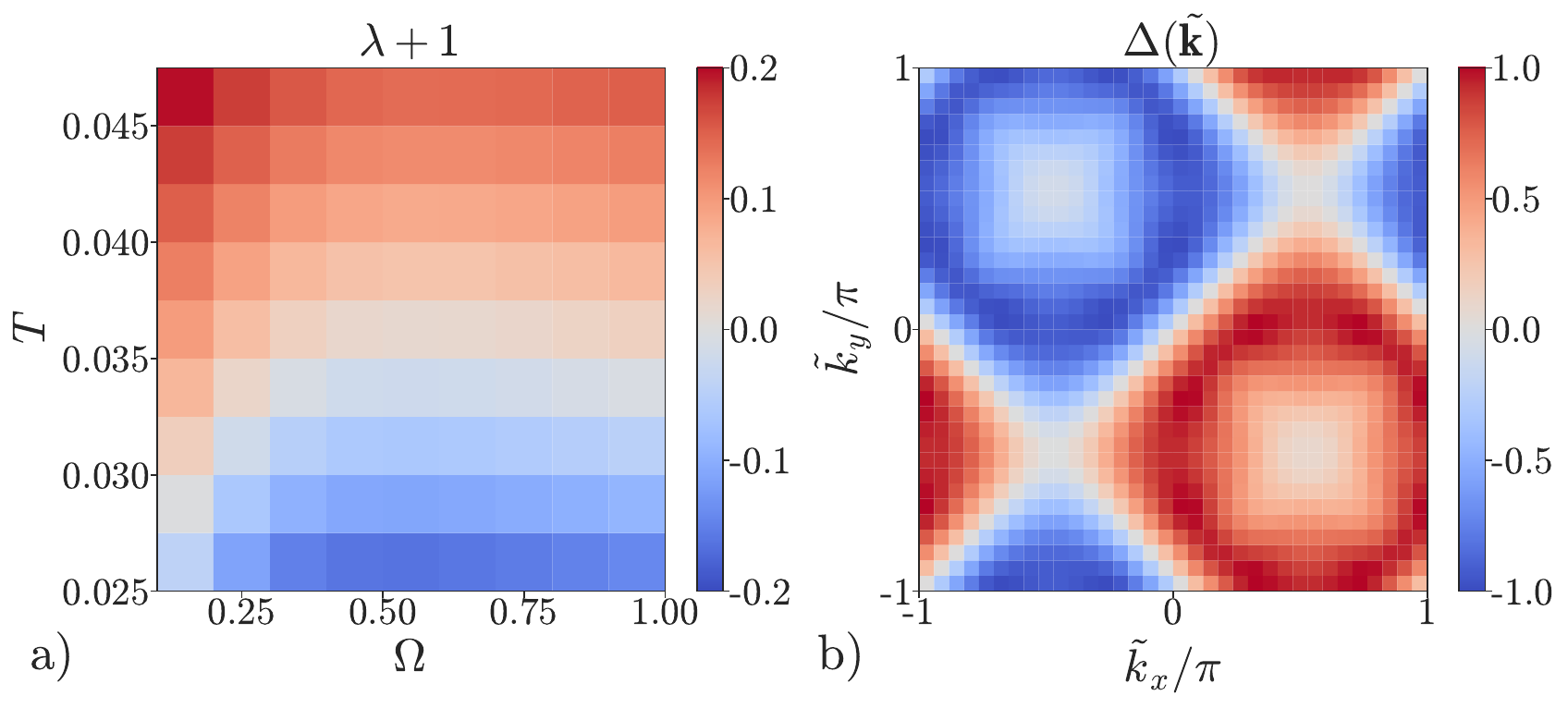}
\caption{(a) $\lambda+1$, with $\lambda$ defined in Eq. \eqref{gapEq}, as obtained from a variational calculation with the ansatz in Eq. \eqref{ansatz}. (b) Corresponding gap function $\Delta(\tilde{\k})$. Results are obtained on a $34\times 34$ pseudo-momentum grid using $m = 0.01$.} \label{fig:lambda}
\end{figure}

To study electron pairing we ignore the empty conduction band, and work exclusively with the valence band which crosses the Fermi energy. We emphasize that this step can only be justified by using the rotating frame, which eliminates inter-band scattering terms of order $U$ \cite{Vasiliou2023,Chubukov1997_2}. We work at a hole doping of $\sim 12\%$. The resulting Fermi surface is shown in Fig. \ref{fig:SeffData} (a). To study superconductivity we sum the usual Cooper ladder diagrams, where each rung consists of both the instantaneous interaction contained in $S_V$, and the interaction generated by tree-level Goldstone mode exchange. The Cooper ladder sum diverges when following equation has a solution with $\lambda = -1$:
\begin{equation}\label{gapEq}
\lambda\hat{\Delta}(i\omega',\tilde{\k}') = T\sum_{i\omega}\frac{1}{N}\sum_{\tilde{\k}}\frac{V(i\omega'-i\omega,\tilde{\k}',\tilde{\k})}{\omega^2+(E_{\tilde{\k}}-\mu)^2} \hat{\Delta}(i\omega,\tilde{\k})\,,
\end{equation}
where $V(i\omega'-i\omega,\tilde{\k},\tilde{\k}')$ scatters a pair of electrons with pseudo-momenta $(\tilde{\k},-\tilde{\k})$ and frequencies $(i\omega,-i\omega)$ to a pair of electrons with $(\tilde{\k}',-\tilde{\k}')$ and $(i\omega',-i\omega')$, due to both the instantaneous interaction and Goldstone mode exchange. The explicit expression for $V$ is given in the supplementary material \cite{supplement}, but let us note here that the frequency dependence of $V$ comes from both the Goldstone mode propagator and the direct coupling of the electrons to $\partial_\tau \phi$ [see Eq. \eqref{SelB}]. In Eq. \eqref{gapEq} we have also introduced the notation $E_{\tilde{\k}} := E_{\tilde{\k},0}$, and used that $E_{\tilde{\k}} = E_{-\tilde{\k}}$. To obtain $\lambda$, we use following ansatz
\begin{equation}\label{ansatz}
\hat{\Delta}(i\omega,\tilde{\k}) = \frac{\Delta(\tilde{\k})}{\omega^2+\Omega^2}\,,
\end{equation}
where $\Delta(-\tilde{\k}) = -\Delta(\tilde{\k})$ as required by fermion antisymmetry, and $\Omega$ is a variational parameter corresponding to an inverse retardation timescale. In Fig. \ref{fig:lambda} (a) we show $\lambda+1$ obtained from a variational calculation with this ansatz as a function of $T$ and $\Omega$, where we have used a thermal Goldstone mass of $m=0.01$ (for details see \cite{supplement}). We see that $\lambda$ reaches $-1$ at a highest temperature of $\sim 0.033$ when $\Omega \approx 0.5$, which is roughly half the Goldstone bandwidth. Due to the variational nature of our calculation we have thus obtained a lower bound for $T_c$ (within the ladder sum approximation) which, when using a representative value of $t=0.3$ eV, corresponds to $\sim 115$ K. In Fig. \ref{fig:lambda} (b) we show the corresponding gap function $\Delta(\tilde{\k})$. It has the important property $\Delta(\tilde{\k}+\Q) = -\Delta(\tilde{\k})$ (recall that the shift symmetry over momentum $\Q$ is a result of the U$(1)$ spin rotation symmetry). We have also calculated $\lambda$ and $\Delta(\tilde{\k})$ using a thermal mass $m=0.1$ and found that in this case $T_c \gtrsim 0.028$, and the gap function remains essentially unchanged \cite{supplement}.

In the spin basis, the gap function is given by $\Delta_{\sigma\sigma'}(\tilde{\k}) = u_{0,\sigma}(\tilde{\k})u_{0,\sigma'}(-\tilde{\k})\Delta(\tilde{\k})$. As the AFM breaks spin rotation symmetry, the gap is generically an admixture of a singlet and a triplet component. To see this, we rewrite the gap function in the crystal momentum basis, where it takes the form $\Delta_{\sigma\sigma'}(\k)=\Delta^S(\k)\sigma^y_{\sigma\sigma'} + \Delta^T_\Q(\k)\sigma^z_{\sigma\sigma'}$ \cite{supplement}. $\Delta^S(\k)$ is a zero-momentum singlet gap, and $\Delta^T_\Q(\k)$ is a momentum-$\Q$ triplet gap \cite{Psaltakis1983,Murakami1998,Kyung2000,Reiss2007,Wang2014}. Written out explicitly, the singlet part is given by
\begin{equation}
\Delta^{S}(\k) = u(-\k-\Q/2)_{0\downarrow}u(\k+\Q/2)_{0\uparrow}\Delta(\k+\Q/2) \,.
\end{equation}
Note that from $\Delta(-\k)=-\Delta(\k)$, $\Delta(\k+\Q)=-\Delta(\k)$ and $u(\k)_{0\downarrow} = u(\k+\Q)_{0\uparrow}$ it follows that $\Delta^S(-\k) = \Delta^S(\k)$ as required. 

\begin{figure}
\includegraphics[scale=0.31]{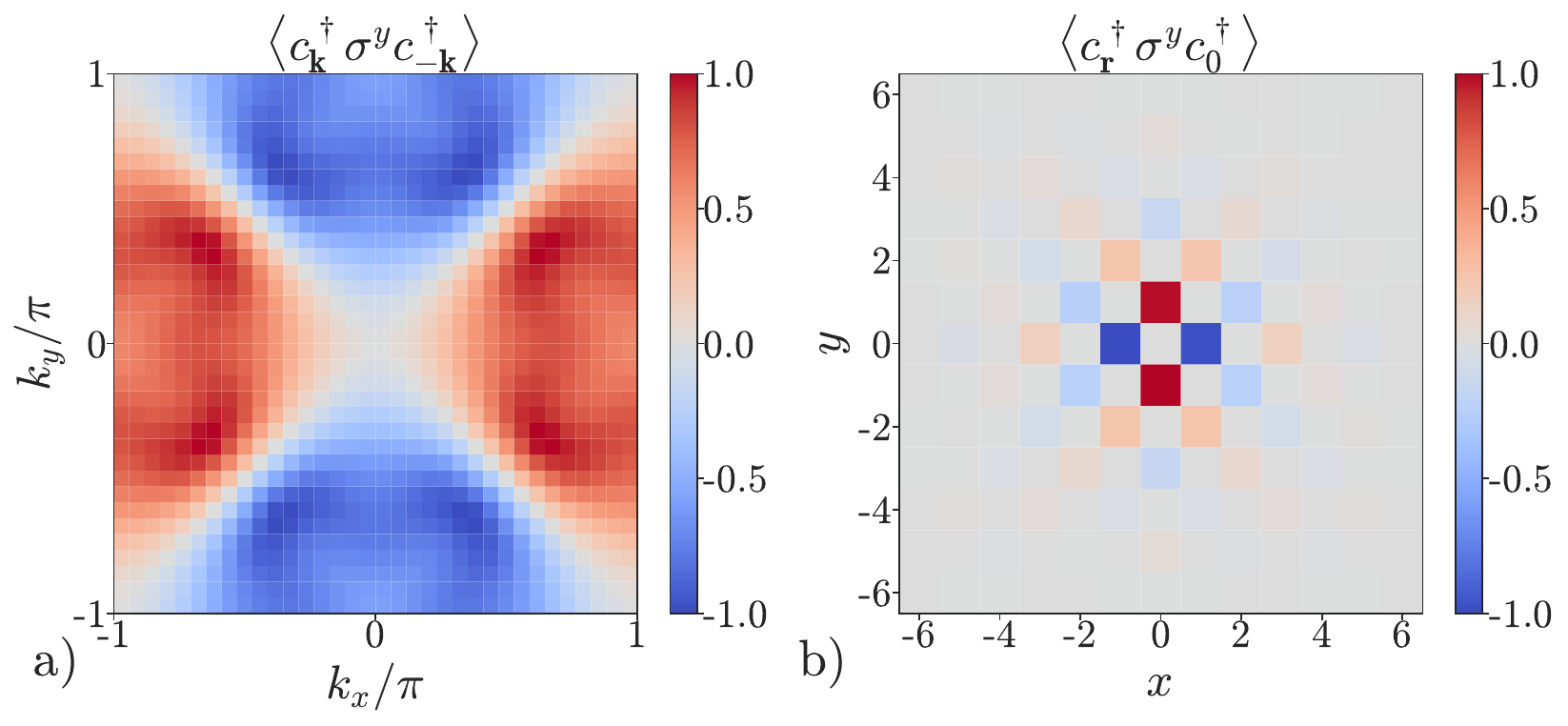}
\caption{(a) Plot of the superconducting order parameter $\langle c^\dagger_{-\k} \sigma^y c^\dagger_{\k}\rangle$ as a function of momentum $\k$. (b) Real-space order parameter/pair wavefunction $\langle c^\dagger_\r\sigma^y c^\dagger_0\rangle$. Results are obtained for a $34\times 34$ system using $m=0.01$. }\label{fig:GapConv}
\end{figure}

Finally, we calculate the superconducting order parameter in terms of the microscopic fermions. We start from the anomalous Green's function
\begin{eqnarray}
\langle \mathcal{T} c^\dagger_{\r,\sigma}(\tau) c^\dagger_{0,\sigma'}(0)\rangle & = & \langle R^*_{\sigma\tilde{\sigma}}(\tau,\r)R^*_{\sigma'\tilde{\sigma}'}(0,0)\bar{\psi}_{\r,\tilde{\sigma}}(\tau)\bar{\psi}_{0,\tilde{\sigma}'}(0)\rangle \nonumber \\
& \approx & \langle R^*_{\sigma\tilde{\sigma}}(\tau,\r) R^*_{\sigma'\tilde{\sigma}'}(0,0)\rangle \times \nonumber \\
 && \langle\bar{\psi}_{\r,\tilde{\sigma}}(\tau)\bar{\psi}_{0,\tilde{\sigma}'}(0) \rangle\,,\label{cdcd}
\end{eqnarray}
where summation over repeated indices is implied, $\mathcal{T}$ is the time-ordering operator, and we have used that the bare electrons are related to the fermion fields $\bar{\psi}_{\r,\sigma}$ via a rotation with $R(\tau,\r)$ as explained above. In the second line we have approximated the correlation function by a product of the correlation functions of the matrix field $R$ and the fermions in the rotating frame. Using $R^* = \sigma^yR\sigma^y$, it immediately follows from the results of Ref. \cite{Borejsza2004} that a saddle-point approximation of the non-linear sigma model describing the dynamics of $R$ leads to following expression for the $R$ two-point function at non-zero temperature:
\begin{equation}
\langle R^*_{ss'}(\tau,\r) R^*_{\tilde{s}\tilde{s}'}(0,0)\rangle = \sigma^y_{\tilde{s}s}\sigma^y_{s'\tilde{s}'}D(\tau,\r)\,,
\end{equation}
where $D(\tau,\r)$ is the Fourier transform of $((i\nu)^2-\omega_\q^2)^{-1}$. As $\langle\bar{\psi}_{\r,\tilde{\sigma}}(\tau)\bar{\psi}_{0,\tilde{\sigma}'}(0) \rangle$ is the Fourier transform of $\Delta_{\tilde{\sigma}\tilde{\sigma}'}(\k)/(\omega^2+\Omega^2)$, we find that the (equal-time) superconducting order parameter is given by
\begin{equation}\label{SCordparam}
\langle c^\dagger_{-\k,\sigma}c^\dagger_{\k,\sigma'}\rangle  =  \sigma^y_{\sigma\sigma'}\frac{1}{N}\sum_\q \frac{n(\omega_\q) - n(-\omega_\q)}{\omega_\q} \Delta^S(\k-\q)\,,
\end{equation}
where $n(\omega)=1/(\exp(\omega/T)-1)$ is the Bose-Einstein distribution function, and a momentum-independent prefactor has been dropped. Note that the contraction of $\Delta_{\sigma\sigma'}(\k)$ with the $R$ two-point function annihilates the momentum-$\Q$ triplet component, such that the superconducting order parameter is translation invariant and pure singlet. This is similar to how a spin-rotation invariant Green's function was obtained in Refs. \cite{Borejsza2004,Vasiliou2023,Wu2018,Scheurer2018,Bonetti2022}. In Fig. \ref{fig:GapConv} we plot the order parameter \eqref{SCordparam} both in momentum and real space. It clearly has a nodal $d$-wave structure, whose origin can be traced back to the single-band nature of the pairing function, which imposes $\Delta(-\tilde{\k}) = -\Delta(\tilde{\k})$, together with the $C_4$ and time-reversal symmetries. From Fig. \ref{fig:GapConv} (b) we also see that the electron pairs are tightly bound together: the pair wavefunction becomes negligibly small when the electrons are separated by more than $\sim 5$ lattice sites. From the same figure we also see that the pair wavefunction is only non-zero when the electrons occupy different sublattices. This is a direct consequence of the U$(1)$ spin rotation symmetry of the AFM, which imprints on the gap function the property $\Delta(\k+\Q)=-\Delta(\k)$. From Eq. \eqref{cdcd} it is also clear that the pair wavefunction will decay faster if the spin moments become more disordered. If we use a thermal mass $m=0.1$ such that the spin correlations become more short-ranged, we find that the pair wavefunction is essentially zero if the electrons are separated by more than one lattice site \cite{supplement}.

\emph{Discussion --} We have shown that with microscopically calculated scattering vertices, finite-temperature remnants of AFM Goldstone modes can mediate a sizeable attractive interaction between electrons which leads to spin-singlet $d_{x^2-y^2}$ superconducting order. The fact that $T_c$ drops only moderately (by $\sim 15\%$) upon increasing the thermal mass of the Goldstone modes by an order of magnitude is consistent with the expectation that pairing is predominantly mediated by short-wavelength and high-energy transverse spin fluctuations, as these couple most strongly to the electrons. The effective theory used in this work is obtained from a simple mean-field+RPA analysis, which makes it unclear how reliable our estimate for $T_c$ is. An interesting route for future research would therefore be to improve our results by using the two-particle self-consistent approach \cite{Vilk1997} applied to broken-symmetry systems \cite{DelRe2023}, or by combining the functional renormalization group with mean-field theory \cite{Reiss2007,Wang2014,Yamase2016,Vilardi2020}.  

In this work we have ignored Landau damping of the Goldstone modes. We expect that this will only quantitatively change our results, and here we only aim to get an order of magnitude estimate for $T_c$. Our analysis does reveal a few distinct universal features of electron pairing in fluctuating itinerant AFM which can be used to test the theory in more controlled numerical studies, or perhaps even in experiment. In particular, our key predictions are: (1) the presence of fluctuating triplet order at momentum $\Q$, (2) a pair wavefunction which decays faster than the spin correlation function, and (3) a purely inter-sublattice pair wavefunction. The latter was also found in the spin-bag approach \cite{Schrieffer1988,Schrieffer1989}, which is not surprising since it is a consequence of the symmetries of the AFM.

The theory presented here is undoubtedly too simplified to capture the rich physics of the underdoped cuprates. For example, it does not take charge stripes, nematicity and pair density-wave orders into account, all of which can intertwine with each other, the AFM, and the uniform superconductor in intricate ways \cite{Fradkin2015,Tranquada2020}. Nevertheless, we believe that our approach can shed light on the role of the higher-temperature pseudogap state as a parent state for superconductivity. Of course, this requires one to assume that the pseudogap state observed in experiment is indeed a fluctuating AFM, which is still a matter of current debate. But at least for the Hubbard model the magnetic pseudogap scenario has received considerable numerical support \cite{Macridin2006,Kyung2006,Gunnarson2015,Wu2017,Wu2018,Wietek2021,Simkovic2022}.

\emph{Ackowledgements -- }The author would like to thank Andrey Chubukov, Walter Metzner and Pietro Bonetti for helpful discussions, and Rafael Fernandes for pointing out Ref. \cite{Psaltakis1983}. This research is supported by the European Research Council (ERC) under the European
Union’s Horizon 2020 research and innovation programme (Grant agreement No. 101076597 - SIESS).

\bibliography{bib}

\clearpage
\newpage
\begin{appendix}
\onecolumngrid
	\begin{center}
		\textbf{\large --- Supplementary Material ---}
	\end{center}

 \section{Appendix A: Definition of $u(\tilde{\k})_{\alpha,\sigma}$ and $S_V$}
 Underlying the effective action in the main text is the mean-field Hamiltonian in the pseudo-momentum basis:
 \begin{equation}
H_{MF} = \sum_{\tilde{\k}} f_{\tilde{\k}}^\dagger h(\tilde{\k})f_{\tilde{\k}}\,,
 \end{equation}
 where $f_{\tilde{\k}} = (f_{\tilde{\k},\uparrow}, f_{\tilde{\k},\downarrow})^T$, with $f^\dagger_{\tilde{\k},\sigma}$ defined in Eq. \eqref{75} in the main text. The precise form of the mean-field Hamiltonian is
 \begin{equation}
h(\tilde{\k}) = \left(\begin{matrix} \varepsilon_{\tilde{\k},\uparrow} & M \\ M & \varepsilon_{\tilde{\k},\downarrow}\end{matrix}\right)\,,
 \end{equation}
where $M \sim 1.93$ is the self-consistently determined (using the parameters for the Hubbard model given in the main text) hybridization between the spin up and down electrons which produces the AFM order, and
\begin{eqnarray}
\varepsilon_{\tilde{\k},\uparrow} & = & \varepsilon(\tilde{\k} - \Q/2)\,, \\
\varepsilon_{\tilde{\k},\downarrow} & = & \varepsilon(\tilde{\k} + \Q/2)\,.
\end{eqnarray}
Here, $\varepsilon(\k) = -2t(\cos k_x + \cos k_y) -2t'(\cos (k_x+k_y) + \cos (k_x-k_y)$ is the dispersion of the Hubbard model in Eq. \eqref{Hubbard}. The mean-field energies are defined as the eigenvalues of $h(\tilde{\k})$, and are given by
\begin{equation}
E_{\tilde{\k},\alpha} = \frac{1}{2}\left(\varepsilon_{\tilde{\k},\uparrow} + \varepsilon_{\tilde{\k},\downarrow} \pm \sqrt{(\varepsilon_{\tilde{\k},\uparrow} - \varepsilon_{\tilde{\k},\downarrow})^2 + 4M^2}\right)\,.
\end{equation}
The states $|u(\tilde{\k})_\alpha\rangle$ are the corresponding eigenstates of $h(\tilde{\k})$. Note that the mean-field Hamiltonian satisfies $h(\tilde{\k}+\Q) = \sigma^x h(\tilde{\k})\sigma^x$, which is a manifestation of the U$(1)$ spin rotation symmetry around the $x$-axis. The eigenstates of $h(\tilde{\k})$ (in a real gauge) thus satisfy  $|u(\tilde{\k}+\Q)_\alpha\rangle = \pm \sigma^x|u(\tilde{\k})_\alpha\rangle$. We partially fix the gauge by requiring these signs to all be positive.

Written out explicitly, the instantaneous interaction in the effective theory used in the main text is given by
\begin{equation}\label{SV}
 S_{V} = \int\mathrm{d}\tau \frac{1}{N}\sum_{\tilde{\q},\tilde{\k},\tilde{\k}'} \left[ \frac{U}{2}\left(\bar{\psi}_{\tilde{\k}-\tilde{\q}}\Lambda_{\tilde{\q}}(\tilde{\k})\psi_{\tilde{\k}}\right)\left( \bar{\psi}_{\tilde{\k}'+\tilde{\q}}\Lambda_{-\tilde{\q}}(\tilde{\k}') \psi_{\tilde{\k}'}\right) - \frac{c}{2aw}\big(\bar{\psi}_{\tilde{\k}} f_{\tilde{\q}}(\tilde{\k})\psi_{\tilde{\k}-\tilde{\q}} \big) \left(\bar{\psi}_{\tilde{\k}'-\tilde{\q}} f_{\tilde{\q}}^{\dagger}(\tilde{\k}')\psi_{\tilde{\k}'} \right)\right]\,.
\end{equation}
The first term in Eq. \eqref{SV} is simply the microscopic Hubbard interaction $\frac{U}{2}\sum_\r:n_\r^2:$ rewritten in the mean-field basis. The unitary basis transformation from the orbital basis to the mean-field basis gives rise to the form factors:
\begin{equation}
\left[\Lambda_{\q}(\k)\right]_{\alpha\beta} = \langle u(\tilde{\k}-\tilde{\q})_\alpha|u(\tilde{\k})_\beta\rangle\, .
\end{equation}
The second term in Eq. \eqref{SV} is obtained by integrating out the field conjugate to the Goldstone field $\phi(\r)$ \cite{Vasiliou2023}. As in the main text $c \approx 0.7$ is the Goldstone mode velocity, $a$ is the lattice constant, and $w \approx 0.66$ is a dimensionless number which is fixed by the requirement that in the rotating frame it should hold that $\lim_{\tilde{\q}\rightarrow 0,\Q}g_{\tilde{\q}} = 0$ \cite{Vasiliou2023}. The matrices $\left[f_{\tilde{\q}}(\tilde{\k})\right]_{\alpha\beta}$ contain the scattering vertices $f_{\tilde{\q},\alpha\beta}(\tilde{\k})$ used in Eq. \eqref{SelB}.

 \section{Appendix B: Definition of $V(i\omega-i\omega',\tilde{\k},\tilde{\k}')$}
The interaction $V(i\omega-i\omega',\tilde{\k},\tilde{\k}')$ which scatters a pair of electrons with frequencies $(i\omega,-i\omega)$ and pseudo-momenta $(\tilde{\k},\tilde{\k})$ to a pair of electrons with $(i\omega',-i\omega')$ and pseudo-momenta $(\tilde{\k}',\tilde{\k}')$ can be written as
\begin{equation}\label{Vom}
V(i\omega-i\omega',\tilde{\k},\tilde{\k}') = V_I(\tilde{\k},\tilde{\k}') + V_G(i\omega-i\omega',\tilde{\k},\tilde{\k}')\,.
\end{equation}
The first term comes from the instantaneous interaction contained in $S_V$ and is given by
\begin{equation}
V_I(\tilde{\k},\tilde{\k}') = U\Lambda_{\tilde{\k}-\tilde{\k}',00} (\tilde{\k})\Lambda_{\tilde{\k}'-\tilde{\k},00}(-\tilde{\k}) - \frac{c}{aw}f_{\tilde{\k}'-\tilde{\k},00}(\tilde{\k}')f_{\tilde{\k}-\tilde{\k}',00}(-\tilde{\k}')\,,
\end{equation}
where we have put the band indices equal to zero because we only consider scattering within the valence band which crosses the Fermi energy. The second term in Eq. \eqref{Vom} is generated by tree-level Goldstone mode exchange and is given by
\begin{equation}
V_G(i\omega-i\omega',\tilde{\k},\tilde{\k}') = \frac{\left[ g_{\tilde{\k}'-\tilde{\k},00}(\tilde{\k}') + (\omega-\omega')f_{\tilde{\k}'-\tilde{\k},00}(\tilde{\k}')\right]\left[ g_{\tilde{\k}-\tilde{\k}',00}(-\tilde{\k}') - (\omega-\omega')f_{\tilde{\k}-\tilde{\k}',00}(-\tilde{\k}')\right]}{(i\omega-i\omega')^2 - \omega_{\tilde{\k}'-\tilde{\k}}^2}\,.
\end{equation}

\begin{figure}
\includegraphics[scale=0.4]{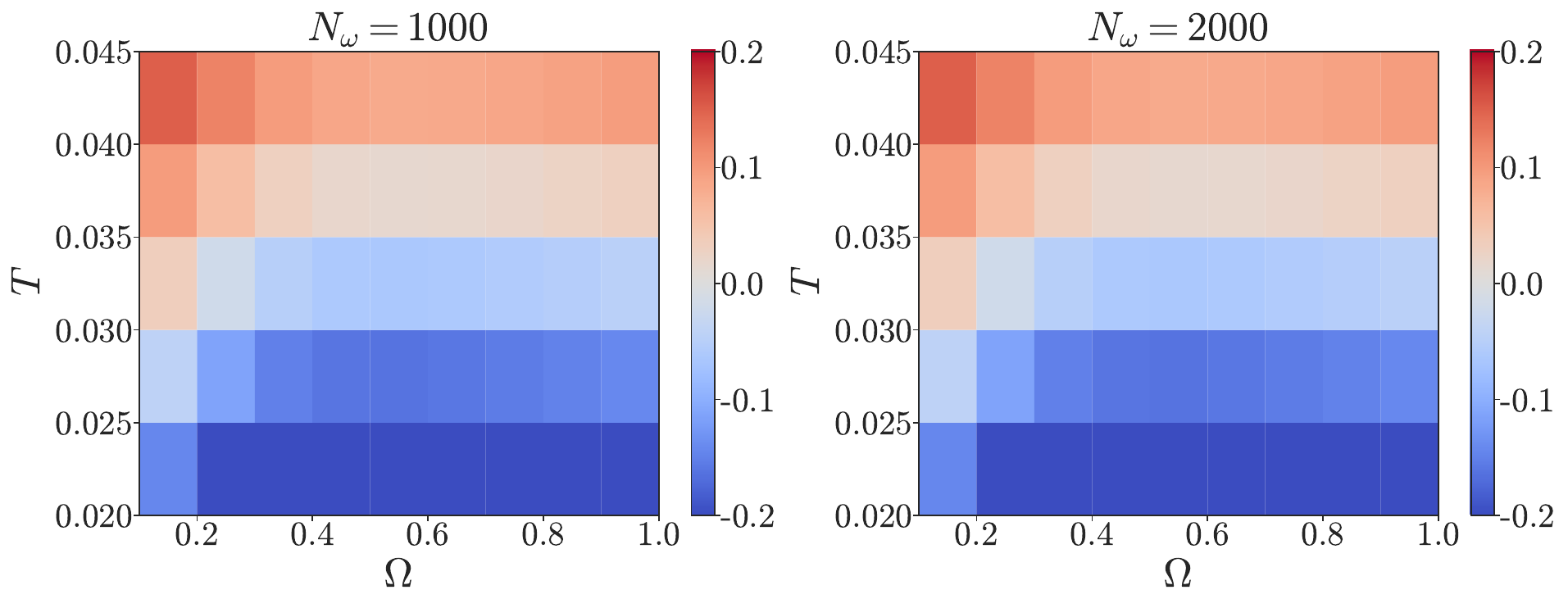}
\caption{Results for $\lambda$ on a $34\times 34$ system using a thermal mass $m=0.01$ and keeping the $N_\omega = 1000$ (left) and $N_\omega=2000$ (right) smallest Matsubara frequencies.}\label{comparison}
\end{figure}

 \section{Appendix C: Variational solution of the gap equation}
 In this appendix we explain in more detail how we solve the gap equation
 \begin{equation}\label{AppgapEq}
\lambda\hat{\Delta}(i\omega',\tilde{\k}') = T\sum_{i\omega}\frac{1}{N}\sum_{\tilde{\k}}\frac{V(i\omega'-i\omega,\tilde{\k}',\tilde{\k})}{\omega^2+(E_{\tilde{\k}}-\mu)^2} \hat{\Delta}(i\omega,\tilde{\k})
\end{equation}
variationally with the ansatz
\begin{equation}\label{Appansatz}
\hat{\Delta}(i\omega,\tilde{\k}) = \frac{\Delta(\tilde{\k})}{\omega^2+\Omega^2}\,.
\end{equation}
As a first step we multiply both sides of Eq. \eqref{AppgapEq} with $\hat{\Delta}^*(i\omega',\tilde{\k}')/(\omega'^2+(E_{\tilde{\k}'}-\mu)^2)$, and sum over both $i\omega'$ and $\tilde{\k}'$. We obtain
\begin{align}\label{Gap1}
& \lambda \sum_{\tilde{\k}} \Delta^*(\tilde{\k}')\Delta(\tilde{\k}') \sum_{i\omega'} \frac{1}{(\omega'^2+\Omega^2)^2}\frac{1}{\omega'^2+(E_{\tilde{\k}'}-\mu)^2} = \\
& \sum_{\tilde{\k},\tilde{\k}'}\Delta^*(\tilde{\k}')\Delta(\tilde{\k}) \frac{T}{N}\sum_{i\omega,i\omega'} \frac{1}{\omega'^2+\Omega^2}\frac{1}{\omega'^2+(E_{\tilde{\k}'}-\mu)^2}V(i\omega'-i\omega,\tilde{\k}',\tilde{\k})\frac{1}{\omega^2+(E_{\tilde{\k}}-\mu)^2} \frac{1}{\omega^2+\Omega^2}\,,\nonumber
\end{align}
where we have filled in the ansatz for the gap function \eqref{Appansatz}. By defining the matrices
\begin{eqnarray}
D(\tilde{\k}',\tilde{\k}) & = & \delta_{\tilde{\k},\tilde{\k}'}\sum_{i\omega'} \frac{1}{(\omega'^2+\Omega^2)^2}\frac{1}{\omega'^2+(E_{\tilde{\k}'}-\mu)^2}\label{sum1}\\
M(\tilde{\k}',\tilde{\k}) & = & \frac{T}{N}\sum_{i\omega,i\omega'} \frac{1}{\omega'^2+\Omega^2}\frac{1}{\omega'^2+(E_{\tilde{\k}'}-\mu)^2}V(i\omega'-i\omega,\tilde{\k}',\tilde{\k})\frac{1}{\omega^2+(E_{\tilde{\k}}-\mu)^2} \frac{1}{\omega^2+\Omega^2}\label{sum2}
\end{eqnarray}
we can bring Eq. \eqref{Gap1} in the form
\begin{equation}
\lambda \langle \Delta|D|\Delta\rangle = \langle \Delta|M|\Delta\rangle\,,
\end{equation}
where $|\Delta\rangle$ is the vector with components $\Delta(\tilde{\k})$. Via following trivial rewriting of this equation
\begin{equation}
\lambda \left(\langle \Delta |D^{1/2}\right)\left(D^{1/2}|\Delta\rangle \right) = \left(\langle \Delta |D^{1/2}\right) D^{-1/2}MD^{-1/2} \left(D^{1/2}|\Delta\rangle \right)
\end{equation}
it becomes clear that the optimal $\lambda$ corresponds to the smallest eigenvalue of $D^{-1/2}MD^{-1/2}$, and the gap function is obtained from the corresponding eigenvector. In practice we evaluate the sums over Matsubara frequencies in Eqs. \eqref{sum1} and \eqref{sum2} with a frequency cutoff. For the results presented in the main text we have kept the 1000 smallest Matsubara frequencies. We have checked that our results are converged upon changing the cutoff. For example, in Fig. \ref{comparison} we compare results obtained by using 1000 and 2000 Matsubara frequencies. We find that the change in $\lambda$ is negligible.

 \section{Appendix D: Singlet and triplet components of the gap function}
 By solving the gap equation we find a gap function which is defined in pseudo-momentum space and in the mean-field band basis. To interpret the gap function physically it is easier to go back to the crystal momentum and spin basis. In this appendix we work out this transformation for the spin-singlet and triplet components of the gap function.

\subsection{Singlet part}
To obtain the singlet component we start from the equations
 \begin{eqnarray}
\langle f^\dagger_{-\tilde{\k},\downarrow} f^\dagger_{\tilde{\k},\uparrow}\rangle & = & \langle c^\dagger_{-\tilde{\k}+\Q/2,\downarrow}c^\dagger_{\tilde{\k}-\Q/2,\uparrow}\rangle \label{eq1}\\
& = & u(-\tilde{\k})_{0\downarrow} u(\tilde{\k})_{0\uparrow}\,\Delta(\tilde{\k})\,,\label{eq2}
 \end{eqnarray}
 where the first line follows from the definition of $f^\dagger_{\tilde{\k},\sigma}$, and the second line follows from the definition of $\Delta(\tilde{\k})$ as the gap function in the mean-field basis. Similarly we also have the equalities
 \begin{eqnarray}
\langle f^\dagger_{-\tilde{\k},\uparrow} f^\dagger_{\tilde{\k},\downarrow}\rangle & = & \langle c^\dagger_{-\tilde{\k}-\Q/2,\uparrow} c^\dagger_{\tilde{\k}+\Q/2,\downarrow}\rangle\label{eq3} \\
 & = & \langle c^\dagger_{-\tilde{\k}+\Q/2 + \Q,\uparrow} c^\dagger_{\tilde{\k}-\Q/2 + \Q,\downarrow}\rangle \label{eq4} \\
 & = & u(-\tilde{\k})_{0\uparrow}u(\tilde{\k})_{0\downarrow}\,\Delta(\tilde{\k})\,. \label{eq5}
 \end{eqnarray}
 Combining Eqs. \eqref{eq1}, \eqref{eq2}, \eqref{eq3}, \eqref{eq4} and \eqref{eq5} we find
 \begin{eqnarray}
\langle c^\dagger_{-\k+\Q/2,\downarrow}c^\dagger_{\k-\Q/2,\uparrow}\rangle & = & u(-\k)_{0\downarrow} u(\k)_{0\uparrow}\,\Delta(\k)\\
\langle c^\dagger_{-\k+\Q/2,\uparrow} c^\dagger_{\k-\Q/2,\downarrow}\rangle & = & u(-\k+\Q)_{0\uparrow} u(\k+\Q)_{0\downarrow} \,\Delta(\k+\Q)
 \end{eqnarray}
A trivial shift in momentum then gives
 \begin{eqnarray}
\langle c^\dagger_{-\k,\downarrow}c^\dagger_{\k,\uparrow}\rangle & = & u(-\k-\Q/2)_{0\downarrow}u(\k+\Q/2)_{0\uparrow} \,\Delta(\k+\Q/2)\label{singlet1}\\
\langle c^\dagger_{-\k,\uparrow}c^\dagger_{\k,\downarrow}\rangle & = & u(-\k+\Q/2)_{0\uparrow}u(\k-\Q/2)_{0\downarrow}\,\Delta(\k-\Q/2)\,.\label{singlet2}
 \end{eqnarray}
 As $\Delta(\k+\Q)=-\Delta(\k)$ and $u(\k+\Q)_{0,\uparrow} = u(\k)_{0,\downarrow}$ the right-hand side Eq. \eqref{singlet1} is equal to minus the right-hand side of Eq. \eqref{singlet2}. Eqs. \eqref{singlet1} and \eqref{singlet2} thus constitute the singlet-component of the gap.

 \subsection{Triplet part}
To obtain the triplet component we follow essentially the same steps as to obtain the singlet component. Starting from the equalities
\begin{eqnarray}
\langle f^\dagger_{-\tilde{\k},\uparrow} f^\dagger_{\tilde{\k},\uparrow}\rangle & = & \langle c^\dagger_{-\tilde{\k}-\Q/2,\uparrow}c^\dagger_{\tilde{\k}-\Q/2,\uparrow}\rangle \\
 & = & u(-\tilde{\k})_{0\uparrow}u(\tilde{\k})_{0\uparrow}\,\Delta(\tilde{\k})
\end{eqnarray}
and
\begin{eqnarray}
\langle f^\dagger_{-\tilde{\k},\downarrow} f^\dagger_{\tilde{\k},\downarrow}\rangle & = & \langle c^\dagger_{-\tilde{\k}+\Q/2,\downarrow} c^\dagger_{\tilde{\k}+\Q/2,\downarrow}\rangle \\
& = & u(-\tilde{\k})_{0\downarrow} u(\tilde{\k})_{0\downarrow}\,\Delta(\tilde{\k})
\end{eqnarray}
we find
\begin{eqnarray}
\langle c^\dagger_{-\k-\Q/2,\uparrow}c^\dagger_{\k-\Q/2,\uparrow}\rangle & = & u(-\k)_{0\uparrow}u(\k)_{0\uparrow}\,\Delta(\k) \\
\langle c^\dagger_{-\k-\Q/2,\downarrow}c^\dagger_{\k-\Q/2,\downarrow}\rangle & = & u(-\k+\Q)_{0\downarrow} u(\k+\Q)_{0\downarrow}\,\Delta(\k+\Q)
\end{eqnarray}
Shifting the momentum then gives
\begin{eqnarray}
\langle c^\dagger_{-\k+\Q,\uparrow} c^\dagger_{\k,\uparrow}\rangle & = & u(-\k -\Q/2)_{0\uparrow} u(\k+\Q/2)_{0\uparrow}\Delta(\k+\Q/2)\\
\langle c^\dagger_{-\k+\Q,\downarrow} c^\dagger_{\k,\downarrow}\rangle & = & u(-\k+\Q/2)_{0\downarrow}u(\k-\Q/2)_{0\downarrow}\,\Delta(\k-\Q/2)\,,
\end{eqnarray}
which corresponds to the triplet component with crystal momentum $\Q$.

\section{Appendix E: Additional numerical results}
In this final appendix we present additional numerical results obtained using a thermal mass $m=0.1$. Other parameters are identical to the ones used in the main text. The system size is $34 \times 34$.

\begin{figure}
\includegraphics[scale=0.4]{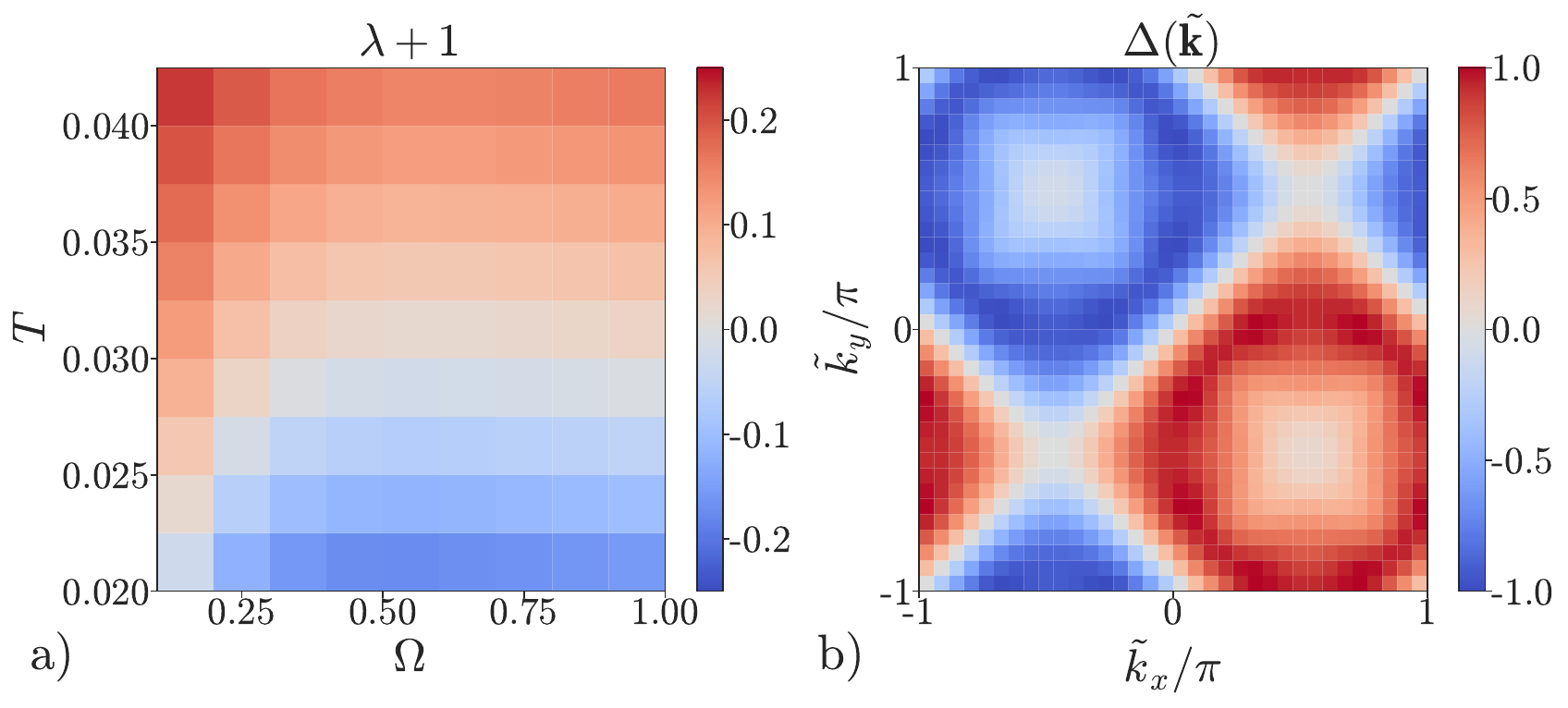}
\caption{(a) $\lambda+1$ as a function of $T$ and $\Omega$ using a thermal mass $m=0.1$. (b) Gap function giving rise to the highest $T_c$ at $\Omega=0.5$. Results are obtained on a $34\times 34$ system.}\label{lambdam01}
\end{figure}

In Fig. \ref{lambdam01} (a) we plot $\lambda+1$ as a function of $T$ and $\Omega$. We see that the highest $T_c \approx 0.028$ occurs for $\Omega \approx 0.5$. The corresponding gap function shown in Fig. \ref{lambdam01} (b) is almost identical to the one obtained in the main text using the smaller thermal mass $m=0.01$.

\begin{figure}
\includegraphics[scale=0.4]{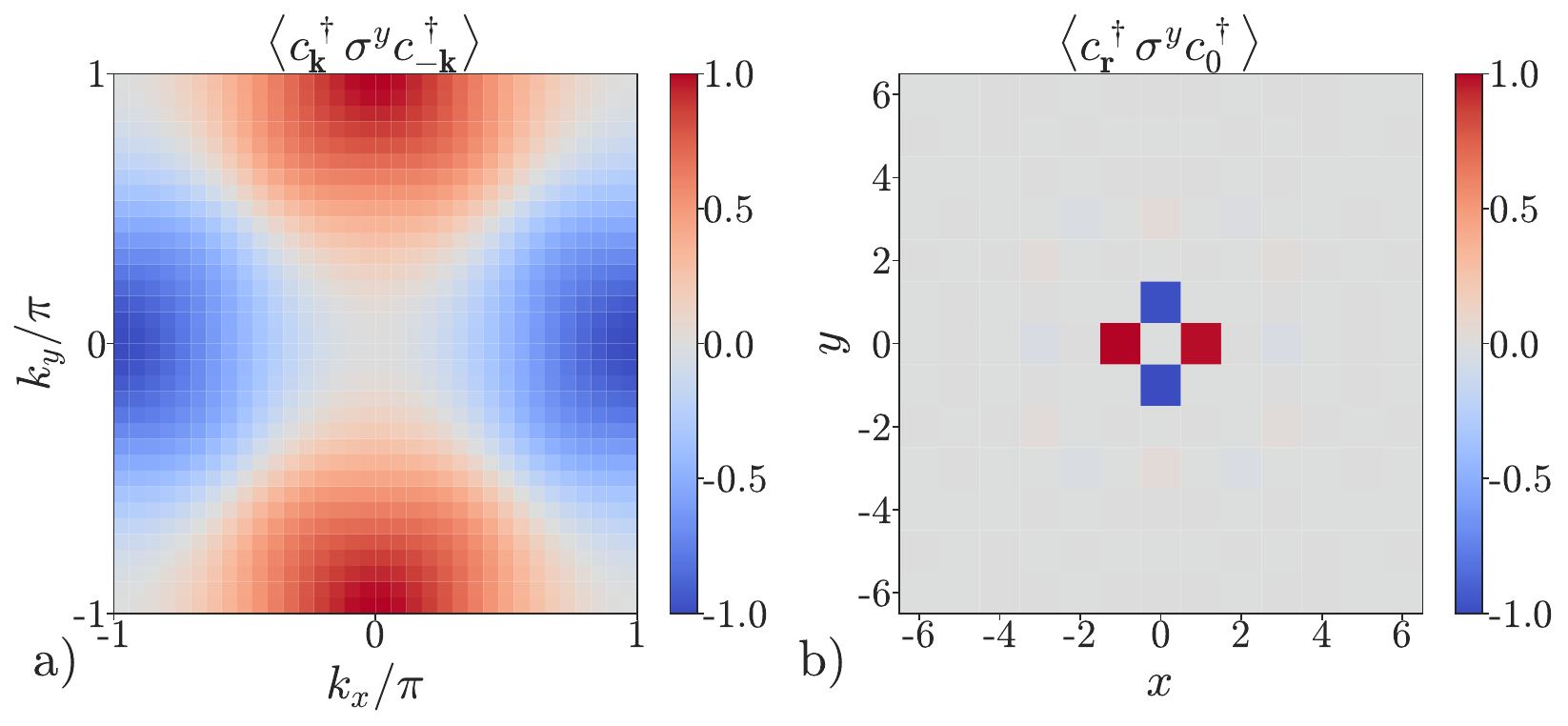}
\caption{Superconducing order parameter and corresponding pair wavefunction obtained with $m=0.1$. (a) Superconducting order parameter as a function of crystal momentum. (b) Pair wavefunction in real-space. Results are obtained on a $34\times 34$ system.}\label{gapm01}
\end{figure}

In Fig. \ref{gapm01} we show the superconducting order parameter defined in Eqs. \eqref{cdcd} and \eqref{SCordparam} in the main text. Compared to the results in the main text obtained with a smaller thermal mass, we see that the order parameter in Fig. \ref{gapm01} is much smoother in momentum space, and hence more short-ranged in real-space. This illustrates the statement made in the main text that our theory predicts a system with shorter-range AFM fluctuations to also have a shorter-range pair wavefunction.

\end{appendix}

\end{document}